\documentstyle[preprint,pre,aps,eqsecnum]{revtex}

\newcommand{\be}{\begin{equation}}
\newcommand{\ee}{\end{equation}}

\newcommand{\ba}{\begin{eqnarray}}
\newcommand{\ea}{\end{eqnarray}}

\newcommand{\binomial}[2]{\left(\begin{array}{c} #1 \\ #2\end{array}\right)}
\newcommand{\qu}[1]{\left[ #1 \right]_q}

\begin{document}

\draft
\preprint{IFUP-TH 34/95}
\draft
\title{Diagrammatic evaluation of conformal weights in the\\
$U_q[SU(2)]$ symmetric Heisenberg chain
}
\author{M. Beccaria}
\address{
Dipartimento di Fisica dell'Universit\`a  and INFN\\
Piazza Torricelli 2, I-56100 Pisa, Italy\\
e-mail: beccaria@hpth4.difi.unipi.it
}
\date{\today}
\maketitle
\begin{abstract}
We consider the $U_q[SU(2)]$ symmetric Heisenberg chain when $q=e^{i\pi/(m+1)}$
and $m$ is
integer.
We consider the cases $m=3$ and $m=5$ which correspond to the Ising and 3-state
Potts
models.
We study the finite size scaling (FSS) of the ground states in
different quantum spin sectors and restricting to highest weights of
type-II representations. We compute the levels by a diagrammatic technique
which needs only the commutation relations of the underlying Temperley-Lieb
algebra.
The results match the FSS predictions which hold for the Bethe levels.
\end{abstract}

\vskip 5mm


\section{Introduction}

The XXZ quantum chain on a finite lattice has a non trivial quantum symmetry
when suitable boundary conditions are considered. Here, we are interested in
the case of a
chain whose symmetry is that of the $U_q[SU(2)]$ quantum
group~\cite{Drinfeld86} with $q$ a root
of unity:
\be
q=e^{i\pi/(m+1)} .
\ee
The chain is massless and the central charge is
\be
c = 1-\frac{6}{m(m+1)}
\ee
and for integer $m$ the the minimal unitary series emerges.

It is very important to understand the structure of the configuration space of
a $N$ sites
chain under the action of the quantum group. For a chain of $1/2$ spins,
$C^{2N}$
decomposes~\cite{Pasquier90} into large indecomposable but
reducible (type-I) representations and irreducible (type-II) representations.

The energy levels can be determined by a Bethe-Ansatz calculation. The
resulting Bethe states
are highest weights with respect to the raising
operator $S^+$. The Kac  conformal weight associated to the
finite size scaling between $U_q[SU(2)]$ spin 0 and spin $j$ ground states is
predicted to be
the entry $h_{1,1+2j}$ in the Kac table.

It is a conjecture that the highest weights of type-II representations
exhaust all the Bethe states with spin $0\le s < m/2$ (see
also~\cite{Koberle95,Pallua95}).

In this letter we study explicitely the finite size scaling of energies and the
corresponding conformal weights in the space of type-II highest weights.
These are associated to the paths in the Bratteli
diagram~\cite{Batchelor91,Hinrichsen94,Arndt95}. In this
path space we can represent the underlying Temperley-Lieb
algebra~\cite{Temperley71} of the XXZ
chain and compute the levels by purely algebraic manipulations.

The states of the path space admits convenient binary representations which
allows
for  simple and compact numerical codes for their manipulation. Relatively
large lattices
may be studied with small computing resources.

The interest of this work is also in showing how the diagrammatic approach
of~\cite{Hinrichsen94} is an effective tool for heuristic investigations, for
instance, we obtain
strong evidence that the above mentioned conjecture holds true.

\section{The XXZ chain and its quantum symmetry}

The hamiltonian is
\be
H = \sum_{i=1}^{N-1} e_i
\ee
where $\{e_1, \cdots, e_{N-1}\}$ are the generators of the Temperley-Lieb
algebra
\ba
e_i^2 &=& x\ e_i \ \ \ \ \ \ x = q+ q^{-1}\\
e_i e_{i\pm 1} e_i &=& e_i \\
\left[e_i, e_j\right] &=& 0\ \ {\rm if}\ \ |i-j|>1
\ea
All the following algebraic manipulations are based on the above relations
only, but
we shall keep in mind the following specific quantum chain representation
\be
e_i = \sigma^x_i\sigma^x_{i+1} + \sigma^y_i\sigma^y_{i+1} +
\frac{q+q^{-1}}{2}\sigma^z_i\sigma^z_{i+1}+\frac{q-q^{-1}}{2}
(\sigma^z_i-\sigma^z_{i+1})
\ee
The $U_q[SU(2)]$ quantum group is generated by
$\{S^\pm, q^{S_z}\}$ with the relations
\ba
\left[S^+, S^-\right] &=& \qu{2S_z} \\
q^{S_z}S^\pm q^{-S_z} &=& q^{\pm 1} S^\pm
\ea
where
\be
\qu{x} = \frac{q^x-q^{-x}}{q-q^{-1}}
\ee
Its spin $1/2$ representation coincides with the classical one ($q=1$) and the
tensored
representations are built as usual by mean of the Hopf-coproduct
\ba
\Delta(q^{\pm S_z}) &=& q^{\pm S_z}\otimes q^{\pm S_z} \\
\Delta(S^\pm) &=& q^{\pm S_z}\otimes S^\pm + S^\pm\otimes q^{\mp S_z}
\ea

The Hamiltonian is exactly invariant under $U_q[SU(2)]$ also at finite $N$
\be
\left[H, S^\pm\right] = \left[H, S_z\right] = 0
\ee

According to~\cite{Batchelor91,Arndt95} we can introduce orthonormal vectors
$V_k$
labelled by
\be
k = (k_0, \cdots, k_N)
\ee
with
\be
\label{con1}
k_0 = 0 \qquad
k_i \ge 0 \qquad
k_{i+1} = k_i \pm 1/2
\ee
If $q$ is a root of unity and $q^p = \pm 1$ then we have the additional
constraint
\be
\label{con2}
k_i \le \frac{p}{2}-1
\ee
related to $\qu{p}$ = 0.
These vectors are in one-to-one correspondence with the highest weights of
type-II representations. The action of the Temperley-Lieb algebra on these
vectors is
\be
e_i\ V_k = \delta_{k_{i-1}, k_{i+1}} \sum_{k^\prime_i = k_{i-1} \pm 1/2}
\sqrt{\frac{\qu{2k_i+1}\qu{2k_i^\prime+1}}{\qu{2k_{i-1}+1}\qu{2k_{i+1}+1}}}
\ V_{k^\prime}
\ee
where
\be
k^\prime = (k_0, \cdots, k_{i-1}, k_i^\prime, k_{i+1}, \cdots , k_N)
\ee
Every weight $k$ corresponds to a representation with spin $k_N$.

\section{Binary representation}

Let us call ${\cal S}_m^{(j)}$ the path subspace with $q=e^{i\pi/(m+1)}$ and
$k_N = j$. We shall study the cases $m=3$ and $m=5$ corresponding to the
Ising model and to the 3-state Potts model. We shall restrict to chains with
an even number of sites.

Every state $s\in {\cal S}_m^{(j)}$ is determined by the
signs $\sigma_i\in\{-1,1\}$ defined by
\be
\sigma_i = 2(k_{i+1} - k_i)
\ee
Let  ${\cal W}_N$ denote the set of $2^{N-1}$ binary words with $N$
bit: every state $s\in {\cal S}_m^{(j)}$ is mapped into a word $w\in{\cal
W}_N$, the $i$-th bit
$b_i$ being $(\sigma_i+1)/2$.
This binary representation is interesting since, in a numerical
study, allows us to write states in terms of integer numbers.
Of course, the $2^{N-1}$ words in ${\cal W}_N$ must satisfy the constraints
Eqs.(\ref{con1},\ref{con2})
imposed over the multilabel $k$. This reduces greatly the dimension of the
state space. Let us turn to some specific examples.

\subsection{$m=3$}

This case corresponds to the Ising model. The structure of the generic path is
well
understood by looking at the following grid which hosts $S_3^{(0)}$ with $N=4$.
\be
\begin{array}{ccccccccccccccccc}
&&&& 1 &&&& 1 &&&& 1 &&&& \\
&&& \nearrow && \searrow && \nearrow && \searrow && \nearrow && \searrow &&& \\
&& 1/2 &&&& 1/2 &&&& 1/2 &&&& 1/2 && \\
& \nearrow && \searrow && \nearrow && \searrow && \nearrow&& \searrow &&
\nearrow && \searrow & \\
0 &&&& 0 &&&& 0 &&&& 0 &&&& 0
\end{array}
\ee
Moreover, $S_3^{(1)}$ is in $1-1$ correspondence with $S_3^{(0)}$ as the
following grid shows.
\be
\begin{array}{ccccccccccccccccc}
&&&& 1 &&&& 1 &&&& 1 &&&& 1\\
&&& \nearrow && \searrow && \nearrow && \searrow && \nearrow && \searrow &&
\nearrow &\\
&& 1/2 &&&& 1/2 &&&& 1/2 &&&& 1/2 && \\
& \nearrow && \searrow && \nearrow && \searrow && \nearrow&& \searrow &&
\nearrow &&& \\
0 &&&& 0 &&&& 0 &&&& 0 &&&&
\end{array}
\ee
We readily see that the relevant signs $\sigma_1$ are only $(N-2)/2$. Thus
\be
\dim S_3^{(0)} = \dim S_3^{(1)} = 2^{N/2-1}
\ee
and every state in $S_3^{(0)}$ or $S_3^{(1)}$ can be mapped in ${\cal
W}_{(N-2)/2}$.

\subsection{$m=5$}

In this case the paths are more complicated due to the weaker constraint
Eq.(\ref{con2}).
An example with $N=10$ could be
\be
\begin{array}{ccccccccccccccccccccc}
&&&&&&&& 2 \\
&&&&&&& \nearrow && \searrow \\
&&&&&& 3/2 &&&& 3/2 &&&& 3/2 \\
&&&&& \nearrow &&&&&& \searrow && \nearrow && \searrow \\
&&&& 1 &&&&&&&& 1 &&&& 1 \\
&&& \nearrow &&&&&&&&&&&&&& \searrow \\
&& 1/2 &&&&&&&&&&&&&&&& 1/2 \\
& \nearrow &&&&&&&&&&&&&&&&&& \searrow \\
0 &&&&&&&&&&&&&&&&&&&& 0 \\
\end{array}
\ee
In the $j=0$ sector every path starts and ends in 0. Therefore the first and
last bits in
${\cal W}_N$ are fixed. The remaining $N-2$ bits are divided equally into
$(N-2)/2$ bits set to 0
and $(N-2)/2$ bits set to 1. The resulting
\be
\binomial{N-2}{(N-2)/2}
\ee
words must satisfy the additional requirement that no spins higher than $2$ do
appear in the
Bratteli diagram.
Defining
\be
\Gamma_k^{(N)} = \binomial{N}{N/2-k} - \binomial{N}{N/2+k+1}
\ee
the dimension of the state space turns out to be
\be
\dim S_5^{0} = \Gamma_0^{(N)} + \sum_{k\ge 1} (\Gamma^{(N)}_{6k} -
\Gamma^{(N)}_{6k-1}) =
\frac{1}{2}(1 + 3^{N/2-1})
\ee
The paths in the sector $j=1$ have only the first bit fixed. At the right end,
$k_N=1$
implies $k_{N-1} = 1/2$ or $k_{N-1} = 3/2$. We find the dimension
\be
\dim S_5^{(1)} = 3^{N/2-1}
\ee
For the paths in the sector $j=2$, two bits are fixed and the total number of
states is
\be
\dim S_5^{(2)} = \frac{1}{2}(3^{N/2-1}-1)
\ee
We utilized $N\le 36$ for the Ising model and $N\le 24$ for the Potts model. At
the largest $N$
we have $\dim S_3^{(0)} = \dim S_3^{(1)} = 131072$, $\dim S_5^{(0)} = 88574$,
$\dim S_5^{(1)} =
177147$ and $\dim S_5^{(2)} = 88573$.

\section{Numerical Results}

Let us define the rescaled energies
\be
\hat{E} = \frac{E}{\displaystyle
2\pi(m+1)\sin\frac{\pi}{m+1}}
\ee
The finite size scaling predictions for the Bethe energy eigenvalues
are~\cite{Pasquier90} the following. Let us consider a chain with $N$ sites and
let
$E_j$ be the ground state energy
in the space of spin $j$ type-II highest weights.
Then
\ba
\label{stress}
\frac{\hat{E}_0}{N} &=& e_\infty + \frac{f_\infty}{N} + \frac{c}{24N^2}
+ \cdots \\
N(\hat{E}_0 - \hat{E}_j) &=& h_{1,1+2j}+ \cdots
\ea
where
\ba
c &=& 1-\frac{6}{m(m+1)} \\
h_{1,1+2j} &=& \frac{j(mj-1)}{m+1}
\ea

We have obtained a numerical estimate of $c$, $h_{1,3}$ and $h_{1,5}$ at $m=3$
and
$m=5$.
As in~\cite{Arndt95} we used the power method in order to determine iteratively
the eigenvector with greates absolute eigenvalue. We stopped the iteration
when the double precision value did not change in the next iteration.
In the determination of $c$ we used the values of $e_\infty$ and $f_\infty$
quoted in~\cite{Alcaraz87}. We used a fitting procedure in order to
eliminate the residual size dependence in Eq.(\ref{stress}).
The energy eigenvalues are shown in Figs (I-II). The extrapolation is
shown in Table I where one can see the very good agreement with conformal
theory.

\section{Conclusions}

In this letter we have shown that the diagrammatic technique prompted out
in~\cite{Hinrichsen94}
is an effective tool. The quantum group symmetry was utilized in~\cite{Arndt95}
in order to build
two-point $U_q[SU(2)]$ scalar operators associated to conformal operators of
the corresponding
critical theory. Here, we have used the quantum symmetry in order to restrict
the analysis of the
spectrum  to a subset of states with well defined symmetry properties. The FSS
of the
energy levels found in the type-II highest weights subspace coincides with that
of the Bethe states
and suggests that all of them are indeed isolated highest weights under the
action of the quantum
group.

\newpage

\section*{Captions}

\noindent{\bf Fig. I  :} Ising model: Energy levels as functions of the lattice
size.

\noindent{\bf Fig. II :} Potts model: Energy levels as functions of the lattice
size.

\begin{table}
\caption{Determination of $c$ and $h_{1,1+2j}$}
\begin{tabular}{ccccccc}
\squeezetable
m & $c$ & $c^{(exact)}$ & $h_{1,3}$ & $h_{1,3}^{(exact)}$ & $h_{1,5}$ &
$h_{1,5}^{(exact)}$\\
\tableline
3  & 0.501(5) & 1/2 &  0.500(5) & 1/2 & -- & -- \\
5  & 0.794(5) & 4/5 &  0.661(5) & 2/3 & 2.993(5) & 3\\
\end{tabular}
\end{table}

\end{document}